%% file: IEEE-TVT.tex
\documentclass[journal]{IEEEtran}

\IEEEoverridecommandlockouts

\ifCLASSINFOpdf
\else
\fi
\usepackage{graphicx}
\usepackage{float} 
\usepackage{textcomp} 
\usepackage[nospace,noadjust]{cite} 
\usepackage{amsmath,amssymb,amsfonts} 
\usepackage{amsthm} 
\usepackage{mathrsfs} 
\usepackage[mathscr]{euscript} 
\usepackage{epstopdf} 
\usepackage{xcolor}
\usepackage{pgfplots}
\pgfplotsset{compat=newest} 
\usepackage{tikz}
\usetikzlibrary{plotmarks}
\usepackage[normalem]{ulem} 
\usepackage{gensymb} 
\usepackage{balance}
\usetikzlibrary{intersections,backgrounds, patterns}
\usepackage{array}
\usepackage[nolist]{acronym} 
\usepgflibrary{arrows}
\usepackage{multirow} 

\usepackage{authblk}

\usepackage[cmintegrals]{newtxmath} 
\usepackage[font=footnotesize]{subcaption}
\usepackage[font=footnotesize]{caption}

\usepackage{algorithm}
\usepackage{algorithmic}

\definecolor{mycolor1}{RGB}{19, 133, 189}
\definecolor{mycolor2}{RGB}{230, 112, 32}
\definecolor{mycolor3}{RGB}{130, 173, 98}
\definecolor{mycolor4}{rgb}{0.49412,0.18431,0.55686}%
\definecolor{myhist}{RGB}{73, 80, 87}
\definecolor{hgreen}{rgb}{0, 0.5, 0}

\pgfplotsset{every axis/.append style={
	scaled x ticks = false,
	label style={font=\footnotesize},
	tick label style={font=\footnotesize},
	tick scale binop=\times
}
}
\pgfkeys{/pgf/number format/.cd,
1000 sep={},
}
\def\plos{\mathcal{P}_\text{LoS}}

\begin{document}
\title{Aerial Vehicles Tracking Using Noncoherent Crowdsourced Wireless Networks}


\author[1]{Hazem~Sallouha\thanks{Hazem Sallouha is funded by PDM Internal Funds - KU Leuven}}
\author[2]{Alessandro~Chiumento}
\author[1]{Sofie~Pollin\vspace*{-0.25cm}}
\affil[1]{Department of Electrical Engineering, KU Leuven, Belgium}
\affil[2]{EEMCS Faculty, University of Twente, The Netherlands}
\affil[ ]{Email: hazem.sallouha@esat.kuleuven.be}

\maketitle
\input{Acro_G2A}
\begin{abstract} 

Air traffic management (ATM) of manned and unmanned aerial vehicles (AVs) relies critically on ubiquitous location tracking. While technologies exist for AVs to broadcast their location periodically and for airports to track and detect AVs, methods to verify the broadcast locations and complement the ATM coverage are urgently needed, addressing anti-spoofing and safe coexistence concerns. In this work, we propose an ATM solution by exploiting noncoherent crowdsourced wireless networks (CWNs) and correcting the inherent clock-synchronization problems present in such non-coordinated sensor networks. While CWNs can provide a great number of measurements for ubiquitous ATM, these are normally obtained from unsynchronized sensors. This article first presents an analysis of the effects of lack of clock synchronization in ATM with CWN and provides solutions based on the presence of few trustworthy sensors in a large non-coordinated network. Secondly, autoregressive-based and long short-term memory (LSTM)-based approaches are investigated to achieve the time synchronization needed for localization of the AVs. Finally, a combination of a multilateration (MLAT) method and a Kalman filter is employed to provide an anti-spoofing tracking solution for AVs. We demonstrate the performance advantages of our framework through a dataset collected by a real-world CWN. Our results show that the proposed framework achieves localization accuracy comparable to that acquired using only GPS-synchronized sensors and outperforms the localization accuracy obtained based on state-of-the-art CWN synchronization methods.

\end{abstract}



\section{Introduction}\label{G2A:intro}

Global air traffic of manned and unmanned \acp{AV} is on a steady rising trend, projecting \acp{UAV} to outgrow \acp{MAV} by several orders of magnitude over the next 20 years \cite{uav_mav}. With \acp{UAV} entering the civil airspace, \ac{ATM} must expand to handle their coexistence with \acp{MAV}, ensuring safe airspace with ubiquitous tracking capabilities \cite{2019reliability,UTM2018uavs,allouch2019,zheng2020,CEDAR,strohmeierKNN}. Unlike the \acp{MAV} safety system, which includes onboard navigation aids, a pilot to intervene, and ground \ac{ATM}, \acp{UAV} safety systems rely primarily on ground \ac{ATM} due to size and power limitations \cite{strohmeierKNN}. In fact, UAV ATM becomes particularly crucial in beyond visual line of sight and autonomous missions.

The essence of ground \ac{ATM} is \acp{AV}' location information typically acquired through \ac{RF}-based surveillance systems. The location information can be obtained independently from the \ac{AV} using primary surveillance radar (PSR), and in an \ac{AV}-dependent manner using \ac{ADS-B} with \ac{AV}'s \ac{GPS}-based location. While existing surveillance systems are sufficient for \acp{MAV}' \ac{ATM}, they are anticipated to face fundamental challenges when dealing with airspace featuring both \acp{MAV} and \acp{UAV} \cite{zheng2020,CEDAR,allouch2019}. First, radar-based solutions have a fixed coverage radius, and the small \ac{UAV} cross-section poses great detection limitations \cite{zheng2020,CEDAR}. Secondly, location-broadcast-based methods, such as \ac{ADS-B}, are expected to suffer from serious reliability and security risks with \acp{UAV} scaling up the number of \acp{AV} in the airspace \cite{2019reliability,allouch2019}. Besides, broadcast methods are broadly vulnerable to spoofing attacks \cite{jansen2018crowd,liu2019}. Alternatively, crowdsourced \ac{ATM} is gaining considerable research focus, as it addresses coverage limitations and promises better resilience against malicious attacks \cite{CEDAR,strohmeierKNN,jansen2018crowd,liu2019}.

Crowdsourced \ac{ATM} exploits the large-scale deployments of \acp{CWN} to capture \acp{AV}' broadcast messages, enabling localization using \ac{TDoA}-based \ac{MLAT} \cite{strohmeierKNN,mlatCOM}. \ac{TDoA}-based \ac{MLAT}, as an independent means of localization, does not require any information from the target \ac{AV}, which facilitates the detection of spoofing attacks. Currently deployed \acp{CWN} rely on off-the-shelf \ac{SDR}-enabled receivers to sense the spectrum and forward the collected data to the network backend via the Internet. While \acp{CWN}' setup guarantees a smooth expansion of the network coverage, it results in a network with noncoherent receivers, consisting of a mix of synchronized and unsynchronized ones \cite{electrosense1}. However, localization using \ac{TDoA}-based \ac{MLAT} requires all receivers involved in the localization process to be synchronized \cite{tdoa_math}.

The tunable \ac{SDR} used in \ac{CWN} receivers enables attractive synchronization opportunities by relying on reference signals from existing wireless networks, e.g., Wi-Fi beacons, LTE synchronization messages, or \ac{ADS-B} messages from trusted aircraft. In this article, we tackle the synchronization problem in noncoherent \acp{CWN} by exploiting the \ac{ADS-B} messages from registered commercial aircraft, along with the limited subset of synchronized receivers, to realize the synchronization needed for \ac{TDoA}-based \ac{MLAT} localization. This promising potential of a crowdsourced-based \ac{AV} localization and tracking, alongside the insisting demand on secure and ubiquitous \ac{ATM}, inspired our work in this article.

\subsection{Related Works}
Recently, the localization and tracking problem of \acp{AV} in general, and \acp{UAV} in particular, has been studied in several articles. The approaches proposed in these articles are broadly categorized into radar systems and localization systems based on \ac{RF} signal emitted from the target \cite{j3hazem}. In \cite{zheng2020}, Zheng \textit{et al.} investigated the detection and localization problem of UAV swarms using a radar method based on the Dechirp-keystone transform and frequency-selective reweighted trace minimization. Guerra \textit{et al.} \cite{2020dynamic} introduced an aerial monostatic dynamic radar network distributed on multiple UAVs to localize and track any malicious UAV in the surrounding area. While both \cite{zheng2020} and \cite{2020dynamic} provided a promising accuracy, it is practically infeasible to deploy such systems for country-wide UAV tracking. On the one hand, it is economically challenging considering the high-cost of radar systems \cite{CEDAR}, and on the other hand, it might be inappropriate or even forbidden to deploy a radar system in urban areas due to the relatively high-power used.

Localization methods based on broadcast \ac{RF} signals from \acp{AV} offer a cost-efficient alternative with the ability to provide country-wide coverage by exploiting the large-scale deployments of \acp{CWN} \cite{electrosense1,2014bringing,strohmeierKNN}. Predominantly, these localization methods employ \ac{MLAT} based on the \ac{RSS} \cite{j2hazem}, \ac{ToA} \cite{abu2016novel}, or \ac{TDoA} \cite{tdoa_math}. In \cite{CEDAR}, Yang \textit{et al.} introduced a cost-efficient crowdsourcing system to detect and localize \acp{UAV}. It has been shown that by exploiting the Wi-Fi beacons broadcast from \acp{UAV}, it is possible to detect and subsequently localize them using \ac{RSS}. However, the considered range was limited to 400\,m; scaling up this range will significantly reduce the localization accuracy, which is inversely proportional to the true distance between the \ac{UAV} and the ground anchor \cite{c2hazem,j3hazem}. Furthermore, \ac{RSS}-based and \ac{ToA}-based methods require knowledge about the transmit power and the transmit time used by the target \ac{AV}, which are not available in the case of a spoofing target \cite{j3hazem}. Such knowledge is not necessary for \ac{TDoA}-based methods, making it a favorable choice for \ac{MLAT} with \ac{CWN} \cite{mlatCOM,seo20193d}. In \cite{seo20193d}, Seo \textit{et al.} proposed a particle filter-based 3D target tracking algorithm with measurement fusion of \ac{TDoA}, frequency difference of arrival (FDoA), and angle of arrival (AoA). Strohmeier \textit{et al.} \cite{strohmeierKNN} introduced a grid-based localization approach for \acp{AV} using the $k$-nearest-neighbor ($k$-NN) algorithm with \ac{TDoA} measurements at a \ac{CWN}. The reported results in both \cite{strohmeierKNN} and \cite{seo20193d} assumed \acp{CWN} with time-synchronized receivers. However, since \acp{CWN} use noncoherent receivers \cite{openskyWB,electrosense1}, this assumption excludes the majority of \ac{CWN} receivers that are not time-synchronized, resulting in sparse \ac{CWN} coverage. For instance, in the OpenSky network, the percentage of time-synchronized receiver sensors is only 15\% of the total number of sensors \cite{openskyWB,2014bringing}. 

The noncoherent and widely-distributed nature of \acp{CWN} restrains them from establishing applications such as cooperative signal decoding and cooperative localization \cite{harris2011let}. An \ac{AR}-based approach, joint with a \ac{KF}, has been adopted in \cite{disty,clktrack} to model the clock behavior in devices with low-precision clocks. While the \ac{AR} process enables promising clock models, the provided results were based on co-located sensors in a single indoor or outdoor environment. In addition, a reference broadcast signal for synchronization was available on-demand. Calvo-Palomino \textit{et al.} \cite{calvo2017} proved that \acp{CWN} could achieve a certain level of synchronization, sufficient for cooperative signal decoding, by using synchronization messages from existing LTE infrastructure. However, this method requires resynchronization with every received message. This requirement is overwhelming for the \ac{CWN} backend, and for the receivers that have limited processing power.

Another promising synchronization reference for noncoherent \acp{CWN} is the \ac{ADS-B} messages from registered commercial aircraft with their trusted broadcast positions. The work presented in \cite{c1hazem} exploits these trusted \ac{ADS-B} messages in conjunction with the limited subset of synchronized receivers to model and subsequently compensate for the clock offset based on an \ac{AR} model. Such clock offset compensation suggests a great potential to achieve the synchronization needed to enable accurate \ac{TDoA}-based \ac{MLAT}. Nevertheless, the results presented in \cite{c1hazem} were only preliminary, lacking details on the clock offset behavior, \ac{AR} model analysis, and the overall localization performance. Moreover, the framework structure and its workflow were not presented.

\subsection{Contribution and Article Structure}
This article addresses \acp{AV}' localization problem using widely distributed noncoherent \acp{CWN}, aiming at large-scale and anti-spoofing crowdsourced-based \ac{ATM}. We consider a real-life \ac{CWN} with receivers classified into 15\% of \acp{GSN} with \ac{GPS}-synchronized clocks and 85\% of \acp{SN} with drifting clocks that are not synchronized. We investigate the clock offset of the \ac{SN} receivers, allowing them to engage in the \ac{TDoA}-based \ac{MLAT} process. We also extend and refine our previous work \cite{c1hazem} by introducing a full localization system design and detailing the workflow for processing both training and test data. Furthermore, we introduce a novel machine-learning-based clock offset modeling, along with comprehensive analyses, concerning the performance of both clock offset modeling and \ac{AV} localization. The main contributions and merits of this article are summarized as follows:

\begin{itemize}
	\item We propose a novel localization framework and a full localization system design, enabling \acp{AV} tracking using a noncoherent \ac{CWN} with a mix of \ac{SN} and \ac{GSN} receivers. The proposed system presents a large-scale \ac{RF}-based \ac{ATM} solution, which can be used to verify the broadcast location of both manned and unmanned vehicles, enabling spoofing attacks detection and tracking.
	
	\item The proposed localization framework investigates two synchronization approaches for noncoherent \acp{CWN}: \ac{AR}-based and \ac{LSTM}-based. We rely on broadcast messages from trusted \acp{AV} and the few available \ac{GSN} to characterize the dynamic el of the \acp{SN}. Subsequently, the resulting model is used to compensate for \acp{SN}' clock offset, enabling target \acp{AV} tracking with \ac{TDoA}-based \ac{MLAT}, joint with a \ac{KF}.
	
	\item Finally, a measurement-based dataset collected from a real-world \ac{CWN} is used to assess the performance of the proposed localization framework. Our results confirm the effectiveness of the proposed framework, minimizing the localization error by around 50\% compared to instantaneous \acp{SN} synchronization using their prior measured offsets.
\end{itemize} 

The rest of the article is organized as follows. In Section \ref{G2A:SM}, we present the system model and the \ac{TDoA} method. Subsequently, the clock offset model is detailed in Section \ref{sec:clkMDL}. Section \ref{nLFW} presents our proposed localization framework with noncoherent \acp{CWN}. Subsequently, we present our experimental results in Section \ref{G2A:RESU}. Finally, Section \ref{G2A:CONC} concludes this article.

\textit{Notation}: Italic letters, simple bold letters, and capital bold letters represent scalars, vectors, and matrices, respectively. $(a_1, a_2, \dots)$ represents a sequence and $[a_1, a_2, \dots]^\text{T}$ represents a column vector, with $[.]^\text{T}$ being the transpose operator. We use $\tilde{x}$, ${x}^-$, and $\hat{x}$ to denote the estimate (or the prediction), the \textit{a priori} estimate, and the \ac{KF}-based estimate, of $x$, respectively.

\section{System Model}\label{G2A:SM}
This section presents the noncoherent \ac{CWN} model considered in this work. Subsequently, it introduces the \ac{TDoA}-based \ac{MLAT} method employed in the proposed framework.

\subsection{Crowdsourced Network Model}\label{CSN}
Consider a ground-based crowdsourced \ac{ATM} network deployed to localize and track \acp{AV}, as illustrated in Fig. \ref{G2Amodel}. We assume that \acp{AV} send periodic \ac{RF} signals which are received by $N$ receivers, denoted by $Rx_1$, $Rx_2$, ..., $Rx_N$ where the $i$th receiver $Rx_i \in \{\text{GSN}\,,\text{SN}\}$ with $i = 1,2,\dots, N$. The receivers register the signal's \acp{ToA}, and subsequently, forward them to a centralized station where the location of the corresponding \ac{AV} is estimated. To develop a suitable localization method, it is crucial first to identify the characteristics of the crowdsourced \ac{ATM} networks. In particular, the communication channel and the receivers' characteristics.

\subsubsection{Communication Channel Characteristics}
\begin{figure}[t]
	\centering	\includegraphics[width=0.48\textwidth]{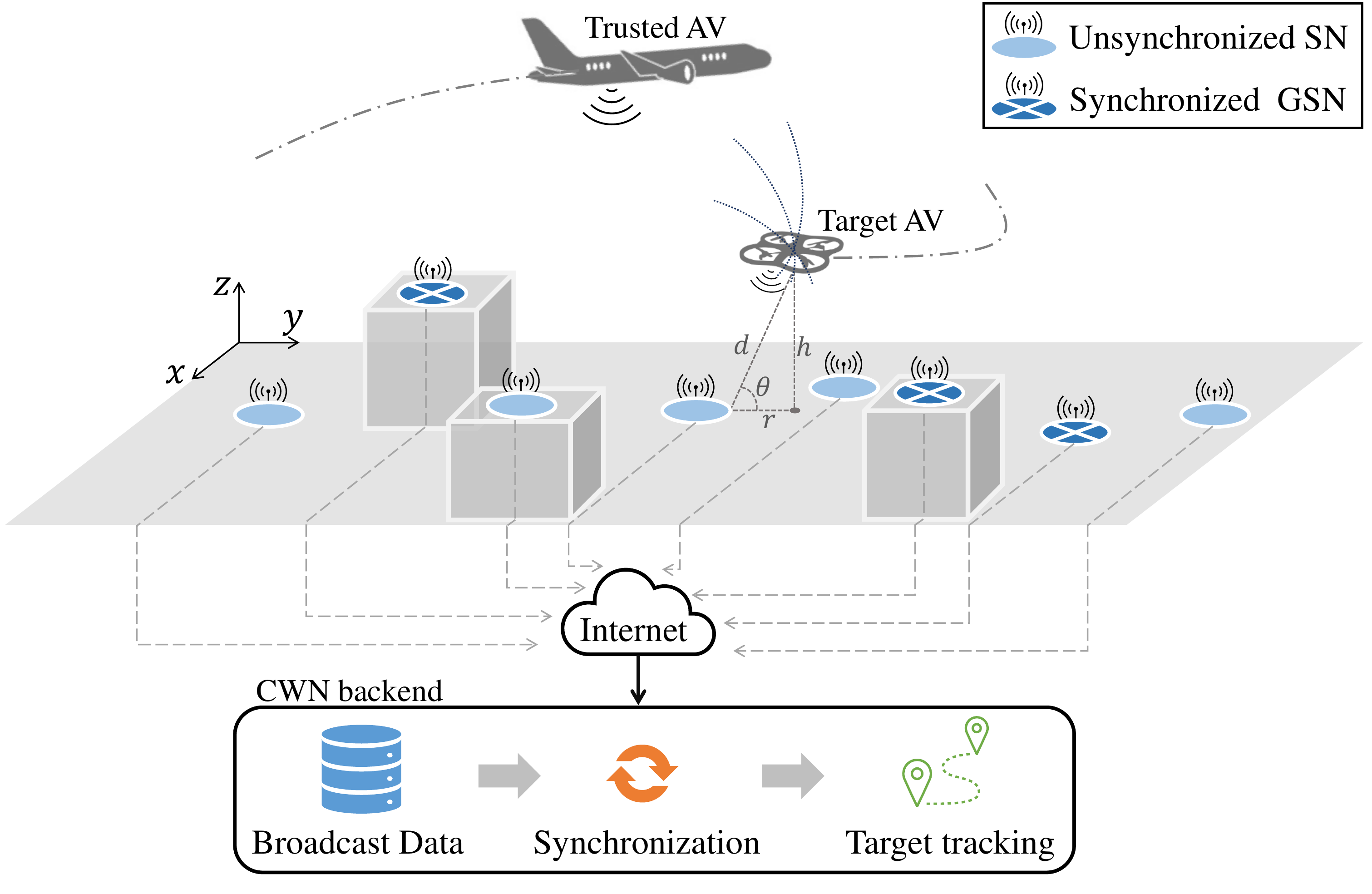} 
	\caption{Target AVs tracking using a noncoherent \ac{CWN} with broadcast messages from trusted AVs used as a synchronization reference.}
	\label{G2Amodel}
\end{figure}

The communication channel between \acp{AV} and ground terminals depends on the propagation environment and the location of the \ac{AV} with respect to the ground terminal. A widely adopted air-to-ground channel model \cite{mozaffari,alzahrani2020} is presented in \cite{hourani}. As reported in this work, the \ac{LoS} and \ac{NLoS} links are considered separately, along with their probabilities of occurrence, and are expressed as

\begin{equation}
	\text{PL}_{n} = 20\log \left(\frac{4\pi fd}{c}\right) + \mu_{n}\,, \,\,\, n \in [\text{LoS}, \text{NLoS}]\,,
	\label{plmodel}	
\end{equation}
where $f$ is the carrier frequency, $c$ is the speed of light, $\mu_{\text{LoS}}$ denotes the mean excessive path loss, and $d$ is the direct distance between the \ac{AV} and the ground terminal, given by $d = \sqrt{h^2 + r^2}$. The probability of having a \ac{LoS} link fundamentally depends on the environment, including density and height of buildings, as well as the elevation angle between the \ac{AV} and the ground terminal. Accordingly, the expression of the \ac{LoS} probability can be written as \cite{hourani,mozaffari}
\begin{equation}
	\plos = \frac{1}{1 + a_o \exp{(-b_o \theta)}}, 
	\label{Prlos}
\end{equation}
where $a_o$ and $b_o$ are environment dependent constants \cite{hourani} and $\theta$ is the elevation angle shown in Fig. \ref{G2Amodel}. The \ac{NLoS} probability is simply $\mathcal{P}_\text{NLoS}$ = $1 - \plos$. Intuitively, for cases where $h >> r$, and $r \neq 0$, we have $\theta \rightarrow 90\degree$, resulting in a $\plos$ that converges to one with roughly free-space path loss model \cite{j3hazem}. For instance, commercial aircraft experience such high \ac{LoS} probability as they fly well-above buildings. While low-altitude \acp{AV} might encounter \ac{NLoS} links, they generally experience a \ac{LoS} probability significantly higher than that of ground-to-ground scenario \cite{j3hazem,hourani,mozaffari}. 

In terms of range, the channel can be classified as a long-range communication channel. In the case of a rotary-wing \acp{AV}, the range varies from a few hundreds of meters to kilometers. This range typically increases by orders of magnitude in the case of fixed-wing \acp{AV}.

\subsubsection{Receivers Characteristics}
Crowdsourced \ac{ATM} networks aim at providing global coverage that is not limited to airports and airfields. To this end, a massive number of receivers with various off-the-shelf hardware are used, e.g., Radarcape, SBS-3, and GRX1090 \cite{2014bringing}. From the TDoA localization perspective, receivers can be categorized as synchronized, namely \ac{GSN}, and unsynchronized, namely \ac{SN}. The \ac{GSN} receivers are usually \ac{GPS} synchronized, meaning that they are constantly resynchronized to compensate for any clock offset. These receivers use a \ac{GPS} disciplined oscillator (GPSDO) as a stable time reference for their local clocks \cite{kaplan2005GPS}. Moreover, \ac{GSN} receivers' timestamps have a rather high resolution. Unsynchronized \ac{SN} receivers, on the other hand, are subject to (sometimes heavy) drifting, and their timestamps have a lower resolution compared to \acp{GSN}. For instance, in the OpenSky network \cite{2014bringing}, \acp{GSN} and \acp{SN} have resolutions of about 40-60\,MHz and 12\,MHz, respectively.

\subsection{MLAT Location Estimation Method}\label{subsec:MLAT}
The \ac{MLAT} process establishes a set of equations relating the \ac{TDoA} measurements, the receivers' positions, and the unknown target position. The approaches used for solving the \ac{MLAT}' set of equations can be classified as statistical, numerical, and algebraic \cite{mlatCOM}. The statistical approaches are highly dependent on the environment's statistical characteristics and are considered open-form algorithms, as they do not provide a closed-form solution \cite{mantilla2012}. Numerical approaches can introduce a closed-form solution; however, a formally driven parameter from the target is required for the numerical approximation \cite{mlatCOM}. Consequently, in the following, we adopt an algebraic approach that uses neither statistical assumptions nor numerical approximations \cite{tdoa_math}, offering an elegant closed-form solution.

Consider a \ac{CWN} employing \ac{MLAT} localization based on \ac{TDoA} measurements among its distributed receivers. The \ac{TDoA} associated with the $i$th receiver, $Rx_i$ and the $j$th receiver, $Rx_j$, is $t_j - t_i$, where $t_i$ and $t_j$ are the \acp{ToA} at $Rx_i$ and $Rx_j$ respectively. Accordingly, one can define the difference in distance $d_{ij}$ as
\begin{align}
	d_{ij} \,\,&{:=}\,\, d_i - d_j \nonumber\\ 
	&= (t_i - t_o)c - (t_j - t_o)c = (t_i - t_j)c \,,
	\label{disDiff}
\end{align}
where $c$ is the speed of light, $t_o$ is the \ac{AV} clock time, and $d_i$ and $d_j$ are the distances from the \ac{AV} to $Rx_i$ and $Rx_j$, respectively. The distance from the \ac{AV} to the $i$th receiver is expressed as
\begin{equation}
	d_i = \sqrt{(x_i - x)^2 + (y_i - y)^2 + (z_i - z)^2}\,,
	\label{disDefn}
\end{equation}
where $(x, y, z)$ is the \ac{AV} position and $(x_i, y_i, z_i)$ is the position of $Rx_i$ with $i = 1, 2, ..., N$. Without loss of generality, the origin of the Cartesian coordinate system is set at $Rx_1$, i.e., $(x_1, y_1, z_1) = (0, 0, 0)$. Consequently, using (\ref{disDefn}), $d_1$ can be written as 
\begin{equation}
	d_1 = \sqrt{x^2 + y^2 + z^2}.
	\label{d1}
\end{equation}
Now, given that $d_{i1} = d_i - d_1$, (\ref{disDefn}) can be rewritten in terms of $d_{i1}$ as
\begin{equation}
	(d_{i1} + d_1)^2 = D_i^2 - 2x_ix-2y_iy - 2z_iz + d_1^2 \,,
	\label{di1Eqn}
\end{equation}
where
\begin{equation}
	D_i^2 = x_i^2 + y_i^2 + z_i^2\,.
	\label{D}
\end{equation}
Expanding (\ref{di1Eqn}), one obtains
\begin{equation}
	-x_ix - y_iy - z_iz = d_{i1}d_1 + \frac{1}{2} \left(d_{i1}^2 - D_i^2\right)\,.
	\label{Redi1Eqn}
\end{equation}
In order to localization a target in 3D, we need $N \geq 4$ \cite{mlatCOM}. Let us define the target position as $\textbf{q} := [x, y, z]^{\text{T}}$, following (\ref{Redi1Eqn}) we can express the problem in matrix format as
\begin{equation}
	\textbf{B} \textbf{q} = d_1 \textbf{m} + \textbf{c} \,,
	\label{MatRedi1Eqn}
\end{equation}
where $\textbf{B}$, $\textbf{m}$, and $\textbf{c}$ are respectively expressed as
\[
\textbf{B}=
\begin{bmatrix}
	x_2 & y_2 & z_2 \\
	x_3 & y_3 & z_3 \\
	\vdots & \vdots & \vdots \\
	x_N & y_N & z_N
\end{bmatrix},\,
\textbf{m}=
\begin{bmatrix}
	-d_{21} \\
	-d_{31} \\
	\vdots\\
	-d_{N1}
\end{bmatrix},\,
\textbf{c}= \frac{1}{2}
\begin{bmatrix}
	D_2^2 - d_{21}^{2} \\[0.4em]
	D_3^2 - d_{31}^{2} \\
	\vdots \\
	D_N^2 - d_{N1}^{2}
\end{bmatrix}
\]
Following \cite{tdoa_math,schau1987passive}, the position of the \ac{AV}, defined as $\tilde{\textbf{q}} := [\tilde{x}, \tilde{y}, \tilde{z}]^{\text{T}}$, can be obtained by solving (\ref{MatRedi1Eqn}) for $\textbf{q}$, which gives
\begin{equation}
	\tilde{\textbf{q}} = \left(\textbf{B}^{\text{T}}\textbf{B}\right)^{-1} \textbf{B}^{\text{T}} \left(d_1 \textbf{m} + \textbf{c} \right).
	\label{posEqn}
\end{equation}
Equation (\ref{posEqn}) contains $d_1$, which is unknown. Substituting (\ref{posEqn}) into (\ref{d1}) yields a quadratic equation in $d_1$. Solving for $d_1$ and substituting the positive root back into (\ref{posEqn}) gives the final solution for $\tilde{\textbf{q}}$. In the case of two positive solutions for $d_1$, we choose the one that lies in the domain of interest, e.g., the one that presents a positive altitude \cite{schau1987passive}. To obtain an accurate position, all receivers need to be synchronized. However, in crowdsourced \ac{ATM} networks, receivers are of two categories: \acp{GSN}, with clocks that are constantly \ac{GPS} synchronized, and \acp{SN}, with clocks that are subject to drifts. In the next section we present the clock modeling and highlight its main design parameters.

\section{Clock Modeling and Noise Analysis}\label{sec:clkMDL}
The broadcast signal from the \ac{AV} is received by multiple receivers, among which some are unsynchronized. Therefore, it is crucial to model the clock's behavior to compensate for any offset. The time difference with the actual time, at the $k$th sample index, is called \textit{time offset} $\eta_{[k]}$, and the instantaneous clock offset rate of change is known as \textit{clock skew} $\lambda_{[k]}$. In fact, the instantaneous time offset depends on the instantaneous clock skew $\lambda_{[k]}$, and the previous clock offset. Accordingly, for a given receiver's clock, the discrete-time clock offset model is expressed as
\begin{equation}
	\eta_{[k]} = \eta_{[k-1]} + \lambda_{[k-1]}\tau_{[k-1]} + \omega_{[k]}\,,
	\label{DclkM}
\end{equation}
where $k$ is the sample index, $\tau_k$ is the $k$th sampling period\footnote{The sampling period $\tau_{[k-1]}$ is the difference in time between the $k$th and the $(k-1)$th samples}, and $\omega_{[k]}$ is a zero-mean normally distributed noise. Following (\ref{DclkM}), we identify two influential parameters in the clock model, the clock skew $\lambda_{[k-1]}$ and the additive noise $\omega_{[k]}$.

\subsection{Clock Skew}
Clock skew results in a clock that runs at a varying speed compared to the actual time. This varying speed is attributed to the oscillator noise, which varies with the supply voltage, age, temperature, and other environmental factors, resulting in a time-varying skew with certain randomness. In general, the oscillator noise is characterized as a nonstationary process; however, the influential parameters' range of change, such as temperature and humidity, varies relatively slowly with time, resulting in a quasi-stationary process \cite{son2020,clktrack}. Therefore, it is of utmost importance to keep track of the quasi-stationarity behavior when modeling the clock skew. In fact, the time-series stationarity characteristics define how often we need to re-calibrate the skew model.

Time-series-based models are used to represent the time-varying skew over an hours-long timescale \cite{clktrack}. Such recursive models must be designed to capture the dynamic as well as the stochastic clock behavior. Once the skew, $\lambda_{[k]}$, is estimated, we exploit the model in (\ref{DclkM}) to predict the clock offset, and subsequently, compensate for it in \ac{TDoA} calculations.

\begin{figure}[t]
	\centering
	\input{clkErrModel.tex}
	\caption{The clock offset histogram of an SN receiver measured with respect to a GSN receiver by relying on common ADS-B messages from trusted AVs, as given in (\ref{offsetM}).}
	\label{clkErr}
\end{figure}
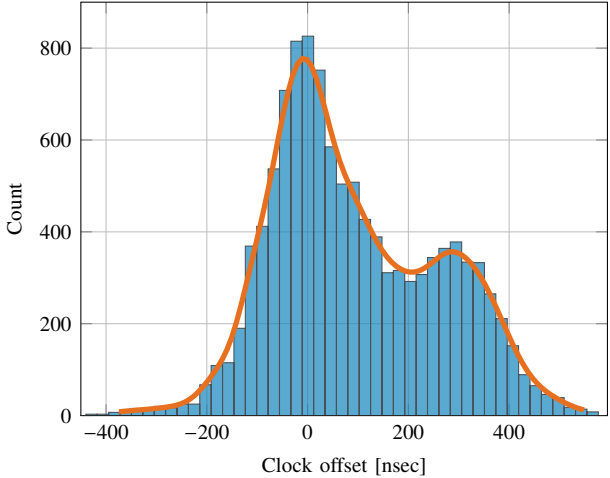
\subsection{Clock Noise and Offset Error}
In addition to the clock skew, the additive offset noise has a noticeable effect on the clock modeling precision. In fact, the vast majority of clock models in the literature assume a normally distributed clock noise \cite{clktrack,disty}. Adding the normally distributed noise to the time-varying clock skew, results in a clock offset with a clear tenancy towards a multimodal distribution. In fact, clocks often drift differently depending on the influential parameter \cite{funck2018synchronous}. For instance, the instantaneous clock skew could vary at different rates with the receiver's printed circuit board temperature, which changes with the processing load. Thus, the same clock could have different clock drift rates on different occasions, resulting in a clock offset with a multimodal distribution.

Figure \ref{clkErr} illustrates the histogram of the clock offset in an \ac{SN} receiver with respect to a \ac{GSN} one. The figure depicts a bimodal error distribution, indicating a skew process with multiple influential parameters \cite{clktrack}. Such non-Gaussian distribution restrains the use of conventional \ac{TDoA} solutions and Bayes filtering algorithms, which are typically designed assuming a Gaussian-type error \cite{KFref1995}. Therefore, after compensating for the offset, it is essential to track the residual clock error distribution to make sure that the non-Gaussian distribution is eliminated and that it follows a normal distribution.

\begin{figure*}[t]
	\centering	\includegraphics[width=0.9\textwidth]{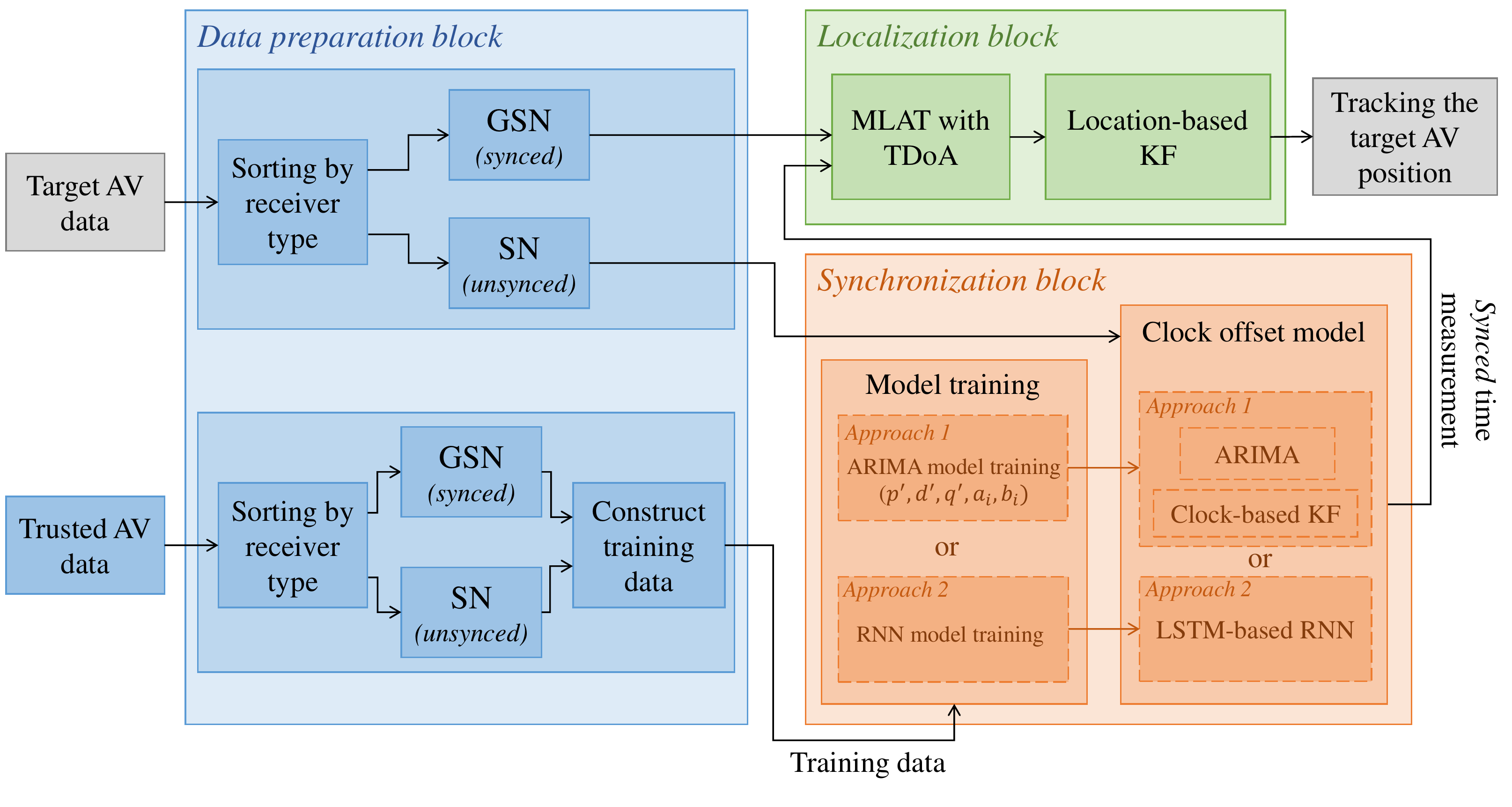} 
	\caption{The block diagram of the proposed AV localization framework illustrating the two clock synchronization approaches considered. Broadcast messages from trusted AVs are exploited to train clock offset models, enabling both GSNs and SNs to engage in the target AV localization.}
	\label{G2Aframework}
\end{figure*}
\section{AV Tracking Using CWNs}\label{nLFW}
In this section, we introduce our localization and tracking framework using noncoherent \acp{CWN}. Figure \ref{G2Aframework} depicts the block diagram of the proposed framework. As shown in the figure, it consists of three main blocks: the data preparation block, the synchronization block, and the localization block. In the following, we detail each block's role in the \acp{AV} tracking process.

\subsection{Data Preparation Block}
The broadcast messages in noncoherent \acp{CWN}, arriving from various \acp{AV}, are captured by two types of sensors: \ac{GSN} and \ac{SN}. This block is mainly responsible for pre-processing and sorting the collected data as well as feeding it to the relevant subsequent block. First, the captured messages' sources must be distinguished as either trusted \acp{AV} or target \acp{AV}. Subsequently, the received messages are sorted based on the receiver's type, namely \ac{GSN} or \ac{SN}. Messages arriving from trusted \acp{AV} are used in constructing the training data as they carry \ac{AV}'s location information, speed, and identification. In particular, the proposed framework utilizes trusted \acp{AV}' broadcast messages mutually received by \acp{GSN} and \acp{SN} to construct the training data for \ac{SN} clock offset modeling. To this end, consider a trusted \ac{AV}'s broadcast message received by a \ac{GSN} receiver and an \ac{SN} receiver. Given the locations of both receivers and the trusted \ac{AV} location, we can express the measured clock offset as
\begin{equation}
	\eta_{[k]} = t_{\text{GSN}[k]} - \Delta_{\text{GSN}[k]} + \Delta_{\text{SN}[k]} - t_{\text{SN}[k]}\,,
	\label{offsetM}
\end{equation}
where $t_{\text{GSN}[k]}$ and $t_{\text{SN}[k]}$ are the measured \acp{ToA} at the \ac{GSN} receiver and \ac{SN} receiver, respectively, $\Delta_{\text{GSN}[k]}$ and $\Delta_{\text{SN}[k]}$ denote the broadcast message's \ac{ToF} from the \ac{AV} to the \ac{GSN} and \ac{SN} receivers, respectively. Note that the \ac{ToF} can be easily measured by dividing the distance between the trusted \ac{AV} and the receiver, as given in (\ref{disDefn}), over the speed of light, $c$. Following (\ref{offsetM}), we construct the training dataset, which is fed to the \textit{synchronization block} where the clock offset model of the corresponding \ac{SN} receiver is estimated. 
Now, messages from target \acp{AV} can be exploited to estimate their locations or verify the broadcast ones. Messages received from the target \ac{AV} by \acp{GSN} are passed directly to the \ac{MLAT} process in the \textit{localization block}, whereas those received by \acp{SN} must first pass through the \textit{synchronization block}, as illustrated in Figure \ref{G2Aframework}.

\subsection{Synchronization Block}
This block models the clock of \acp{SN} and compensates for any clock offset. It takes the \ac{ToA} registered at a given \ac{SN}, from the target \ac{AV} broadcast, as an input. Besides, it also exploits the \acp{ToA} recorded at both \acp{SN} and \acp{GSN}, from trusted \acp{AV}, as training references. Since commercial \acp{AV} send \ac{ADS-B} messages every second, one might intuitively suggest using the prior measured offset to compensate for the subsequent clock offset. Despite the wide availability of \ac{ADS-B} messages, less than 35\% of these messages are received by more than one receiver due to the low reliability of \ac{ADS-B} messages in urban areas \cite{2019reliability}. In addition, receiving the \ac{ADS-B} message by multiple receivers does not necessarily mean having a \ac{GSN} one to act as a reference in measuring the clock offset of other \acp{SN}. As a result, consecutive clock offset measurements are separated by relatively long time gaps, e.g., several minutes. Modeling the clock offset enables predicting the clock offset during these time gaps, which guarantees the continuity of the localization service. The proposed \textit{synchronization block} investigates two approaches to model the clock offset and synchronize \acp{SN}' \ac{ToA} measurements: \ac{AR} and \ac{LSTM}-based \ac{RNN}.

\subsubsection{Autoregressive-Based Synchronization}

When using the recursive model in (\ref{DclkM}) to model the clock offset, one needs to measure the skew, $\lambda_{[k]}$. The clock skew is time-varying; however, for a given measurement duration, it can be modeled using an \ac{AR} process \cite{clktrack}. In particular, \ac{ARIMA} considers the dynamic as well as the stochastic behavior of the clock, and is confirmed suitable for modeling clock skew \cite{disty,clktrack}. Consider an \ac{ARIMA} model, with orders $p'$,$d'$, and $q'$; the skew can be represented by

\begin{equation}
	\underbrace{\left( 1 - \sum_{i=1}^{p'} a_i L^i \right)}_{\text{autoregressive}} \,\, \underbrace{ \vphantom{\left( 1 - \sum_{i=1}^{p'}\right)} \left(1-L \right)^{d'}}_{\text{integrated}}\, \lambda_{[k]} = \underbrace{\left( 1 + \sum_{i=1}^{q'} b_i L^i  \right) \xi_{[k]}}_{\text{moving average}}\,,
	\label{clkSK}
\end{equation}
where $L$ is the lag operator\footnote{The lag operator raised to the power $i$ implies that $L^i \lambda_{[k]} = \lambda_{[k-i]}$}, $\xi_{[k]}$ denotes a zero-mean Gaussian noise, $a_i$'s and $b_i$'s  are the \ac{AR} and the moving average coefficients, respectively. The model in  (\ref{clkSK}) represents an \ac{ARIMA}($p',d',q'$) process where $p'$, $d'$, and $q'$ are non-negative integers representing the order of the autoregressive model, the degree of differencing, and the order of the moving average, respectively. \ac{ARIMA} models are particularly attractive for skew modeling because of their ability to handle any possible non-stationarity in the data \cite{disty,hyndman}. The differencing degree $d'$ is set in order to achieve stationarity in the time-series dataset used for model training and testing. Applying the KPSS test for stationarity \cite{kwiatkowski1992testing} reveals that, based on the receiver, a differencing degree of zero or one $d' = [0, 1]$ is sufficient to guarantee a stationary clock offset for all \ac{SN} receivers of the \ac{CWN} considered in this work. The orders of the autoregressive $p'$ and the moving average $q'$ can be identified using the Box-Jenkins methodology \cite{box2015time}, which relies on the \ac{ACF} and \ac{PACF} of the time-series data (\textit{cf.} Subsection \ref{SyncAnalysis}). The identified state-space formulation of the ARIMA model, results from substituting (\ref{clkSK}) in (\ref{DclkM}), allows a KF formulation \cite{disty}. Therefore, as presented in \cite{disty} and \cite{clktrack}, a \ac{KF} is typically used in conjugation with the model presented in (\ref{DclkM}), to track the time-varying skew and offset, further mitigating any residual error of the ARIMA model.

\begin{table}[t]
	\caption{RNN Model Architecture}
	\centering
		\begin{tabular}{ | c || l | l |}
			\hline
			\textbf{Layer index} & \hspace{2em}\textbf{Type} &  \hspace{3.5em}\textbf{Details}\\
			\hline
			\hline
			1 & Sequence input & - \\
			\hline
			\multirow{3}{*}{2}& \multirow{3}{*}{LSTM} & Units: 10\\
			\cline{3-3}
			& & State activation: \textit{tanh}\\
			\cline{3-3}
			& & Gate activation: \textit{sigmoid} \\
			\hline
			3  & Fully connected & Units: 5 \\ 
			\hline
			4 & Dropout & 0.2\\
			\hline
			5 & Fully connected & Units: 1 \\
			\hline
			6 & Regression output & Mean squared error\\
			\hline
	\end{tabular}
	\label{lstmStuc}
\end{table}

\subsubsection{LSTM-Based Synchronization}
The \ac{LSTM} network, a type of \ac{RNN}, is known for its effective handling of long-term dependencies in time-series data \cite{lstm1997}. The \ac{LSTM} network has feedback connections, enabling predictions based on time-dependent characteristics. Since the clock offset is by default a time-series process, \ac{LSTM} network can learn its behavior over time, and subsequently, make predictions for future values.

The proposed \ac{RNN} consists, first, from an \ac{LSTM} layer with 30 hidden units for the time-series-based clock offset learning. Subsequently, the LSTM layer is followed a fully connected layer, a dropout layer and another fully connected layer. Finally, a regression output layer is presented, delivering the output of the clock offset learning model. The Adam optimizer \cite{adam}, a first-order gradient-based optimizer, with a learning rate of 0.01 is used for clock model training. Table \ref{lstmStuc} details the architecture of the \ac{LSTM}-based \ac{RNN} model used in this work.

Clock offset models provide predictions for the clock behavior. In particular, we train the models so that we can predict $\eta_{[k]}$ based on $\eta_{[k-1]}$. However, the nondeterministic nature of the clock skew prevents a particular clock model from precisely tracking its behavior, resulting in a model's residual error. Consider the \ac{AR} and the \ac{LSTM} clock offset models, let $\tilde{\eta}_{[k]}$ denote the $k$-th sample model-based clock offset prediction; the residual error of the corresponding model is then expressed as
\begin{equation}
	e_{r[k]} = \eta_{[k]} - \tilde{\eta}_{[k]} \,.
	\label{EsdErr}
\end{equation}
where $\eta_{[k]}$ represents the true clock offset.

\subsection{Localization Block}

The localization block consists of the \ac{MLAT} localization method, introduced in Subsection \ref{subsec:MLAT}, followed by a \ac{KF}. This combination of the \ac{MLAT} and \ac{KF} enables tracking of the target \ac{AV} position over time. The movement of the \ac{AV} is modeled as a dynamic system. Given the periodic transmissions from an \ac{AV}, the state vector at the $k$th transmission is written as
\begin{equation}
	\textbf{s}_{[k]} = [x_{[k]},\,y_{[k]},\,z_{[k]},\,\dot{x}_{[k]},\,\dot{y}_{[k]},\,\dot{z}_{[k]}]^\text{T} \in  \mathbb{R}^6 ,
	\label{stateV}
\end{equation}
where the sequence $(x_{[k]}, y_{[k]}, z_{[k]})$ denotes the position of the \ac{AV}, and $(\dot{x}_{[k]}, \dot{y}_{[k]}, \dot{z}_{[k]})$ represents its velocity. Moreover, the input of this dynamic system is represented by the \ac{AV}'s acceleration which, at the $k$th transmission, is given by
\begin{equation}
	\textbf{u}_{[k]} = [\ddot{x}_{[k]},\,\ddot{y}_{[k]},\,\ddot{z}_{[k]}]^\text{T} \in  \mathbb{R}^3 ,
	\label{stateIn}
\end{equation}
where the sequence $(\ddot{x}_{[k]}, \ddot{y}_{[k]}, \ddot{z}_{[k]})$ represents the acceleration of the \ac{AV}. Now, assuming four receivers are available, one can use (\ref{posEqn}) to calculate the \ac{AV}'s position using \ac{TDoA}. Consequently, we express the $k$th measurement vector $\tilde{\textbf{q}}_{[k]}$, which is the calculated position from (\ref{posEqn}), as 
\begin{equation}
	\tilde{\textbf{q}}_{[k]} := [\tilde{x}_{[k]},\tilde{y}_{[k]},\tilde{z}_{[k]}]^{\text{T}} \in \mathbb{R}^3.
	\label{measureV}
\end{equation} 
Finally, the dynamic system can be written as
\begin{align}
	\textbf{s}_{[k]} &= \boldsymbol{\Phi}_{[k-1]}\,\textbf{s}_{[k-1]} + \boldsymbol{\beta}_{[k-1]}\,\textbf{u}_{[k-1]} + \textbf{w}_{[k-1]}, \label{Dmodel1}\\
	\tilde{\textbf{q}}_{[k]} &= \textbf{H}\,\textbf{s}_{[k]} + \textbf{v}_{[k]},
	\label{Dmodel2}
\end{align}
where $\boldsymbol{\Phi}_{[k]} \in \mathbb{R}^{6\times6}$ is the state transition matrix, $\boldsymbol{\beta}_{[k]}\in \mathbb{R}^{6\times 3}$ is the input matrix, $\textbf{u}_{[k]} \in \mathbb{R}^{3}$ is the model input vector, and $\textbf{H}\in \mathbb{R}^{3\times 6}$ represents the measurement matrix. Moreover, in (\ref{Dmodel1}) and (\ref{Dmodel2}), vectors $\textbf{w}_{[k]}$ and $\textbf{v}_{[k]}$ represent the model and measurement noise, respectively. They are assumed to be independent and normally distributed, i.e.,
\begin{align}
	\textbf{w}_{[k]} &\sim \mathcal{N}(0,\textbf{Q}_{[k]})\,\,\,\text{with}\,\,\, \textbf{Q}_{[k]} = \mathbb{E}[\textbf{w}_{[k]}\textbf{w}_{[k]}^{^{\text{T}}}],
	\nonumber\\
	\textbf{v}_{[k]} &\sim \mathcal{N}(0,\textbf{R}_{[k]})\,\,\,\text{with}\,\,\, \textbf{R}_{[k]} = \mathbb{E}[\textbf{v}_{[k]}\textbf{v}_{[k]}^{^{\text{T}}}],\nonumber
\end{align}
where $\mathbb{E}[.]$ denotes the expected value. Now, the \ac{KF} formulation including the predictions and the updates is expressed as \cite{KFref1995}\\ \\
\textbf{\textit{Predict}:}
\begin{subequations}
	\begin{flalign}
		&\hat{\textbf{s}}_{[k]}^- = \boldsymbol{\Phi}_{[k-1]}\,\hat{\textbf{s}}_{[k-1]} + \boldsymbol{\beta_{[k-1]}}\,\textbf{u}_{[k-1]} & 
		\label{Predict1}\\
		&\textbf{P}_{[k]}^- = \boldsymbol{\Phi}_{[k-1]} \textbf{P}_{[k-1]} \boldsymbol{\Phi}_{[k-1]} + \textbf{Q}_{[k-1]} &
		\label{Predict2}
	\end{flalign}
\end{subequations}
\textbf{\textit{Update}:}
\begin{subequations}
	\begin{flalign}
		&\textbf{G}_{[k]} = \textbf{P}_{[k]}^- \textbf{H}^\text{T} ( \textbf{H}\textbf{P}_{[k]}^- \textbf{H} + \textbf{R}_{[k]} )^{-1} & 
		\label{update1}\\
		&\hat{\textbf{s}}_{[k]} = \hat{\textbf{s}}_{[k]}^- + {\textbf{G}}_{[k]} ( \tilde{\textbf{q}}_{[k]} - \textbf{H} \hat{\textbf{s}}_{[k]}^- ) & 
		\label{update2}\\
		&\textbf{P}_{[k]} = ( \textbf{I} - \textbf{G}_{[k]}\textbf{H} )\textbf{P}_{[k]}^- &
		\label{update3}
	\end{flalign}
\end{subequations}
where $\textbf{I}$ is an identity matrix, $\textbf{P}_{[k]} \in \mathbb{R}^{6\times6}$ is the state error covariance matrix, and $\textbf{G}_{[k]} \in \mathbb{R}^{6\times3}$ is the \ac{KF} gain. The \ac{KF} presented in equations (\ref{Predict1})-(\ref{update3}) is able to overcome the Gaussian noise from the \ac{TDoA} calculated position due to the clock offset noise.

\begin{algorithm}[t]
	\caption{Tracking \acp{AV} using the proposed framework}
	\label{alg1}
	\begin{algorithmic}[1]
		\STATE \textbf{Input:} AV, AVDynamicModel, clockARIMAmodel, receivers: $Rx_1$, $Rx_2$, $Rx_3$, \dots
		\STATE \textbf{Output:} Estimated location track of the AV ($\hat{\textbf{s}}_{[k]}$)
		\item[]   
		\WHILE{True}
		\STATE \textit{Msg} $\leftarrow$ newBroadcastMessage(AV)
		\STATE $currentReceivers$ $\leftarrow$ receivers.hasReceived(\textit{Msg})
		\STATE $N$ $\leftarrow$ count($currentReceivers$)
		\STATE \textit{ToA} $\leftarrow$ $currentReceivers$.getToA(\textit{Msg})
		\item[]--------------------- \ac{AR}-based approach ---------------------
		\FOR{$\forall$ $currentReceivers$ \textbf{where} type $==$ SN}
		\item[] \#\,using (\ref{DclkM}), (\ref{clkSK})
		\STATE \textit{ToA} $\leftarrow$ KF$_1$(\textit{ToA},\,clockARIMAmodel) \hspace{3.65cm}
		\ENDFOR
		\item[]------------------- LSTM-based approach -------------------
		\FOR{$\forall$ $currentReceivers$ \textbf{where} type $==$ SN}
		\STATE \textit{ToA} $\leftarrow$ predict(\textit{ToA},\,lstmRNNmodel) \hspace{3.65cm}
		\ENDFOR
		\item[]
		\STATE \textit{TDoAs} $\leftarrow$ calculateTDoAs(\textit{ToA})
		\IF{$N \geq 4$}
		\STATE $k$ $\leftarrow$ getAVtransmissionIndex(\textit{Msg})
		\item[] \#\,using (\ref{posEqn})
		\STATE $\tilde{\textbf{q}}_{[k]}$ $\leftarrow$ MLAT(\textit{TDoAs},\,$currentReceivers$.location)
		\item[] \#\,using (\ref{Predict1})-(\ref{update3})
		\STATE $\hat{\textbf{s}}_{[k]}$ $\leftarrow$ KF$_2$($\tilde{\textbf{q}}_{[k]}$,\,AVDynamicModel)
		\ENDIF
		\ENDWHILE
	\end{algorithmic}
\end{algorithm}

In order to assess the performance of the proposed framework, we use the localization error as a performance metric. Considering a 3D Cartesian coordinate system, we defined the localization error as
\begin{equation}
	\mathcal{E} = \sqrt{(x - \hat{x})^2 + (y - \hat{y})^2 + ((z - \hat{z})/10)^2}\,,
	\label{errTDoA}
\end{equation}
where $(x, y, z)$ is the true position of the \ac{AV} and $(\hat{x}, \hat{y},\hat{z})$ is the \ac{KF}-based estimated one. In (\ref{errTDoA}), we give less weight to altitude estimation, since $x$-$y$ position is typically more critical than altitude estimation in tracking applications.

\begin{table*}[t]
	\caption{Representation of the Dataset Structure}
	\label{osData}
	\centering
	\begin{tabular}{cccccc}
		\firsthline \hline
		\multicolumn{5}{c}{\hspace{2.8cm}$Rx_1$} & \multicolumn{1}{c}{\hspace{0cm}$Rx_N$}\\
		\cline{4-6}
		{Time [sec]}  & avID  & $N$ & [ID$_1$, location$_1$, ToA$_1$, type$_1$]  & \ldots & [ID$_N$, location$_N$, ToA$_N$, type$_N$]\\
		\hline
		T = 0.5 &  630 & 5 & [12, (46.6810,7.6653,10), 121.., GSN] & \ldots & [22, (46.0810,7.2653,8), 123.., SN]   \\[1ex]
		T = 1.2 & 1033 & 4 & [63, (52.3564,4.9522,12), 113.., SN] & \ldots & [24, (44.0810,6.2653,20), 173.., SN]   \\[1ex]
		\vdots  & \vdots  & \vdots    & \vdots	& \ldots & \vdots \\ [1ex]
		\lasthline
	\end{tabular}
\end{table*}

\subsection{Workflow and Algorithm Summary}

A summary of the proposed framework is presented in Algorithm \ref{alg1}. The algorithm works as follows. With each broadcast from \acp{AV}, the algorithm collects the \ac{ToA} from all available receivers (\textit{currentReceivers}). Subsequently, it checks the type of each receiver. For each unsynchronized receiver (i.e., type == SN), it passes its data to the synchronization process, with the corresponding approach being either \ac{AR}-based or \ac{LSTM}-based. Once all \acp{ToA} are synchronized, they are used to calculate \acp{TDoA}, which are subsequently used to estimate the location ($\tilde{\textbf{q}}_{[k]}$) by employing \ac{MLAT}. The estimated location is then processed using a \ac{KF} (i.e., KF$_2$) with the \ac{AV} dynamic model to obtain the final location estimate ($\hat{\textbf{s}}_{[k]}$).

\section{Experimental Results}\label{G2A:RESU}
In this section, we assess the performance of the proposed method when applied on a dataset collected by the OpenSky Network \cite{openskyWB}. The dataset consists of \ac{ADS-B} messages sent from commercial aircraft and received by a total of 523 receivers, uniformly distributed around Europe, with 15\% of them are \acp{GSN} and the rest are \ac{SN} receivers. The dataset is stored in different files, where each file represents a one-hour recording of \ac{ADS-B} messages received by all 523 receivers. The stored data includes the actual time at the server, the aircraft ID, the number of available receivers for each message, and the receivers' information. The receiver information contains its ID, location, \ac{ToA}, and type. Table \ref{osData} depicts a visualization of the dataset. In addition, the true locations of the aircraft are also available. The true position is used to calculate the localization error from (\ref{errTDoA}). It is worth noting that although the considered dataset is based on \ac{MAV}, the performance analysis is also valid for other \ac{AV}. This due to the fact that the \ac{LoS} probability $\plos$ given in (\ref{Prlos}) approaches one at altitudes above the average buildings height \cite{j3hazem}, leading to a steady \ac{TDoA} performance above this altitude. Moreover, similar to the \ac{ADS-B} technology in \acp{MAV}, the majority of \acp{UAV} use Wi-Fi for communication \cite{CEDAR}, allowing \acp{CWN} to capture their Wi-Fi beacons \cite{minucci2020}. 

\begin{figure}[t]
	\centering
	\input{ACF.tex}
	\caption{The sample ACF of the measured clock offset with one degree of differencing ($d' = 1$).}
	\label{figACF}
\end{figure}
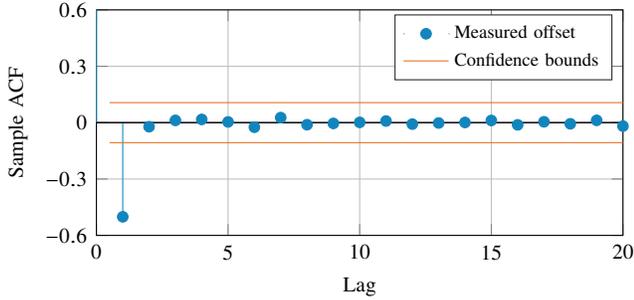 
\begin{figure}[t]
	\centering
	\input{PACF.tex}
	\caption{The sample PACF of the measured clock offset with one degree of differencing ($d' = 1$).}
	\label{figPACF}
\end{figure}
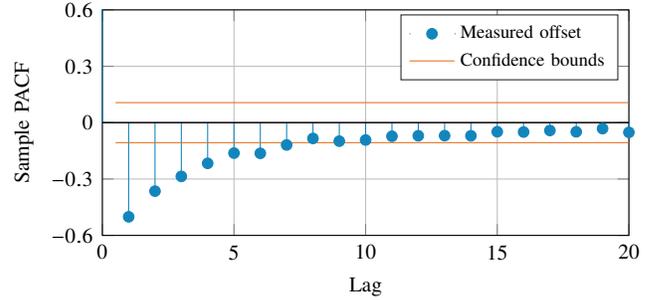

To construct the training and the test datasets, we consider a one-hour recording of \ac{ADS-B} messages mutually received by four nodes, which corresponds to a total of approximately 12700 messages. First, we filtered out messages from eight aircraft and used them as a test dataset, representing the target \acp{AV}. The reaming dataset is assumed to be collected from trusted \acp{AV}, which are used to measure the clock offset of \acp{SN} using equation (\ref{offsetM}). Subsequently, we train both the \ac{LSTM}-based \ac{RNN} and the \ac{ARIMA} models based on messages within a window of 20 minutes. The resulting models are then used to predict the clock offset for the subsequent time instants at which messages from the target \acp{AV} arrive. The models are retrained again after 20 minutes with an updated training data of the same size. This 20 minutes retraining period is chosen empirically to 1) guarantee a sufficient amount of common new messages (e.g., $>$ 100 messages over a span of 20 minutes) between the \ac{SN} and a \ac{GSN} for effectual retraining, 2) bound the predictions' cumulative error, and 3) make sure that the clock offset measurements used in model training and testing represent a stationary process. The time required to train the clock offset models varies based on the size of the available training messages. For the considered \ac{ARIMA} model, once $p'$, $d'$, and $q'$ are identified, the training time ranges from 0.5 to 3 seconds, whereas for the \ac{LSTM}-based model, the training time ranges from 30 to 60 seconds\footnote{A desktop computer without a dedicated Graphics processing unit (GPU) is used for training the models considered in this work.}. In the following, we first analyze the proposed clock synchronization approaches. Subsequently, the localization performance of the proposed framework is investigated.

\subsection{Synchronization Analysis}\label{SyncAnalysis}

To identify the order of the \ac{ARIMA} model, we adopt the Box-Jenkins methodology \cite{box2015time}, which recommends choosing the autoregressive order $p'$ and the moving average order $q'$ based on the number of lags that have values above the confidence bounds of the \ac{PACF} and \ac{ACF}, respectively. Figure \ref{figACF} and Figure \ref{figPACF}, respectively, present the \ac{ACF} and \ac{PACF} of an SN's clock offset with 90\% concordance bounds. In both figures, we introduce one degree of differencing to make sure that the considered clock offset is a stationary process. The negative correlation in the figures is attributed to this differencing. Following Figure \ref{figACF}, we set $q' = 1$. Similarly, based on Figure \ref{figPACF}, we infer an \ac{AR} order $p' = 6$. Putting all together, the resulting model is ARIMA(6,1,1). Here, it is worth noting that for the \acp{SN} considered in this work, where no differencing is needed to achieve stationarity, similar orders $q'$ and $p'$ are observed from the corresponding \ac{ACF} and \ac{PACF} plots, resulting in an ARIMA(6,0,1) model.

\begin{figure}[t]
	\centering
	\input{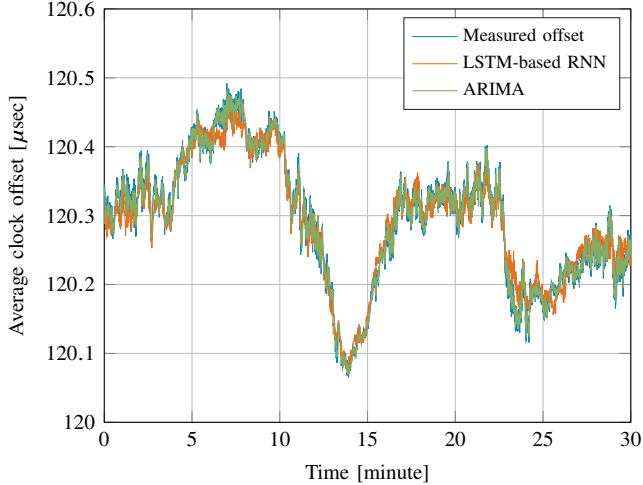}
	\caption{A compression of clock modeling using LSTM-based RNN and ARIMA-based approaches. Both approaches achieve rather accurate modeling.}
	\label{clkmodelComp}
\end{figure}

Figure \ref{clkmodelComp} presents the clock offset modeling when using LSTM and \ac{ARIMA} over 30 minutes period. As shown in the figure, both approaches provide a rather accurate offset modeling that ranges in the margin of 400 nanoseconds. As illustrated in the figure, both approaches confirmed their ability to cope with the clock offset, despite its nonlinearity and randomness.

\begin{figure}[t]
	\centering
	\input{fittingResidualhistn_2.tex}
	\vspace{-0.4cm}
	\caption{Probability plot comparing the distribution of the clock offset residuals to the normal distribution.}
	\label{fittingResidualHist}
\end{figure}
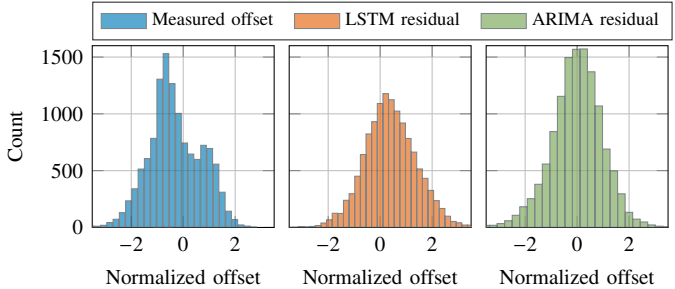

As stated in Section \ref{sec:clkMDL}, it is crucial to make sure that the clock model's residual error follows a normal distribution. This normality is needed for the subsequent localization block. Figure \ref{fittingResidualHist} presents the histogram plots for an \ac{SN} receiver's clock offset. The figure depicts the bimodal distribution of the normalized offset before any compensation. As shown in the figure, employing \ac{AR} and \ac{LSTM} models to predict and compensate for the clock offset results in residual errors ($e_{r[k]}$'s) with a bell-shaped normal distribution. Accordingly, it can be confirmed that both \ac{ARIMA} and \ac{LSTM} clock models are able to successfully eliminate the bimodal distribution, leading to normally distributed residual errors. Such normally distributed error can be handled by the \ac{KF} in the subsequent localization block.

\subsection{Localization and Tracking}

\begin{figure}[t]
	\centering
	\input{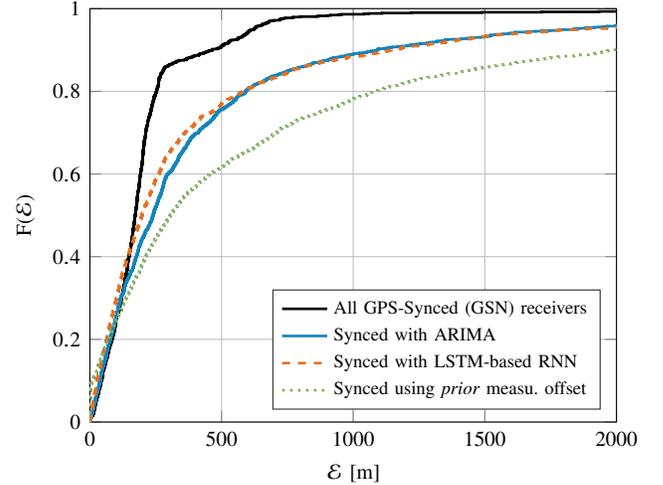}
	\caption{The empirical CDF of the localization error with for both LSTM-based RNN and ARIMA-based synchronization approaches.}
	\label{G2Acdf2}
\end{figure}

Figure \ref{G2Acdf2} demonstrate the empirical \ac{CDF} of the localization error using the proposed framework. The figure represents the localization error based on 1250 received broadcasts, which correspond to eight aircraft broadcasting messages over 40 minutes period. In this scenario, we consider four receivers, two of which are unsynchronized, i.e., two \acp{GSN} and two \acp{SN}. Figure \ref{G2Acdf2} compares between the two synchronization approaches considered. In addition, the figure presents a scenario in which all receivers are GPS-synchronized, i.e., GSN receivers. Analyzing the figure, several observations can be drawn.

\begin{enumerate}
	
	\item First, the figure shows that the considered ARIMA and LSTM-based synchronization models roughly provide a matching performance in terms of the localization error probability. However, the figure exhibits a slight performance gain for the LSTM-based approach over the ARIMA-based approach. This gain is attributed to the structure of LSTM-based \ac{RNN} that enables it to learn both long-term and short-term correlations and discard all the irrelevant ones in between \cite{lstm1997}. 
	
	\item Furthermore, Figure \ref{G2Acdf2} proves the significant localization gain obtained by using the proposed framework. The proposed framework brings the localization performance nearly 50\% closer to GSN receivers' performance compared to relying on the \textit{prior} measured offset for synchronization without any clock modeling, similar to the synchronization method adopted in \cite{calvo2017}. For instance, using the proposed framework's synchronization approaches, e.g., ARIMA or LSTM, brings the 80\% error probability upper-bound from 1100\,m using \textit{prior} offset down to 600\,m.
	
	\item Lastly, Figure \ref{G2Acdf2} emphasizes that even with all receivers being GSP-synchronized, it is still possible to experience localization error in the range of hundreds of meters. This shows that the achieved accuracy has other limiting factors in addition to the clock synchronization. Examples of these factors are the receivers' orientation with respect to the target \cite{strohmeierKNN,tdoa_math} and the packets' time-tagging mechanize that may varies from one receiver to another \cite{zhu2020airsync}.
	
\end{enumerate}

\begin{figure}[t]
	\centering
	\input{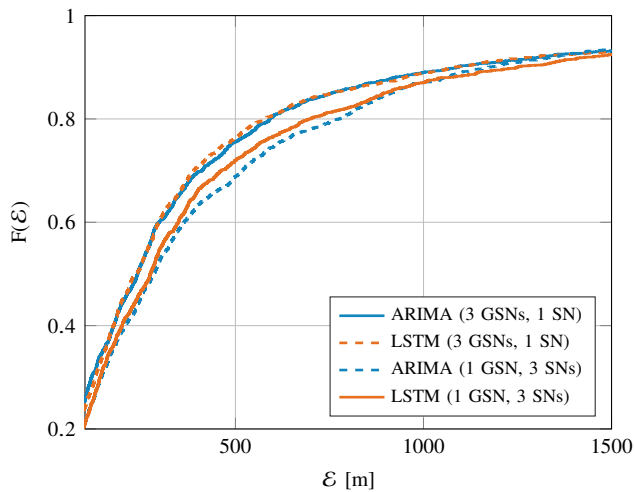}
	\caption{The empirical CDF of the localization error with different numbers of GSNs and SNs.}
	\label{numSNcdf}
\end{figure}

\begin{figure}[t]
	\centering
	\includegraphics[width=0.45\textwidth]{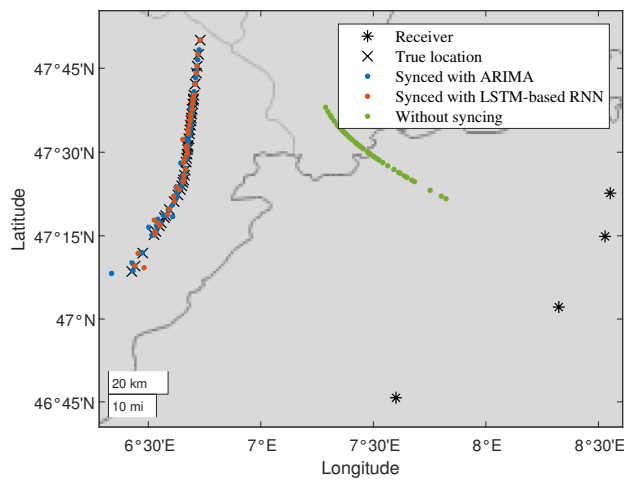}
	\caption{The localization gains obtained by employing the proposed framework in a CWN with AV's trajectory parallel to the receivers' positions.}
	\label{Res1}
\end{figure}

\begin{figure}[t]
	\centering
	\includegraphics[width=0.45\textwidth]{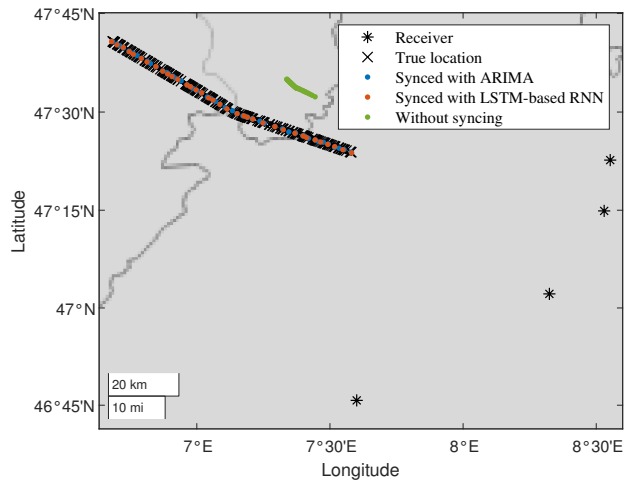}
	\caption{The localization gains obtained by employing the proposed framework in a CWN with AV's trajectory perpendicular to the receivers' positions.}
	\label{Res2}
\end{figure}

Since the proposed framework relies on GSNs to synchronize SNs as well as to take part in the localization process, it is important to explore the impact of the proportion of \acp{GSN} in the CWN. The effect of the proportion of \acp{GSN} and \acp{SN} employed to localize the same target \acp{AV} from Figure \ref{G2Acdf2} is presented in Figure \ref{numSNcdf}. As the figure illustrates, the scenario with three \acp{GSN} and one \ac{SN} outperforms the scenario where one GSN and three SNs are used. By comparing both scenarios, we confirm the positive impact of having more GSNs involved in the target localization. Nonetheless, the localization performance difference between the two scenarios is rather small, proving the considerable impact of the proposed synchronization methods. For instance, with LSTM-based synchronization, Figure \ref{numSNcdf} shows a roughly 3\% difference in performance between the scenarios with one GSN and three GSNs. In fact, this difference in performance with the proportion of employed \acp{GSN} and \acp{SN} is notably less pronounced compared to the effect of receivers' orientation with respect to the target \ac{AV} (\textit{cf.} Figure \ref{Res1} and Figure \ref{Res2}).

The localization gains obtained with the proposed framework are also depicted in Figure \ref{Res1} and Figure \ref{Res2}. The figures exhibit both synchronization approaches, along with the case without synchronization. The \ac{MLAT} cases without synchronization, presented in Figure \ref{Res1} and Figure \ref{Res2}, illustrate the significant influence of the considered receivers' clock offset on deteriorating the localization performance before employing the proposed framework. Figure \ref{Res1} presents an aircraft trajectory roughly parallel to the receiver's locations. The corresponding localization errors are summarized in Table \ref{AVres1}. As shown in the table, using the ARIMA model to compensate for the clock offset decreases the error by orders of magnitude. Comparable performance is also achieved with LSTM-based synchronization. Furthermore, Table \ref{AVres1} highlights the effect of the location-based \ac{KF} (i.e., KF$_2$) on the localization performance, refining the performance of the \ac{TDoA}-based \ac{MLAT} process.

The performance of \ac{MLAT}-based localization is considerably influenced by the target's position relative to the receivers' positions \cite{strohmeierKNN,tdoa_math}. In Figure \ref{Res2}, we present another scenario, with the same set of receivers, where the \ac{AV}'s trajectory is perpendicular to the curve connecting the receivers' locations. As shown in Table \ref{AVres2}, better performance is achieved with the same set of receivers and the same clock model. In particular, using the proposed framework with LSTM-based synchronization, an average localization error of 155\,m is achieved. Similarly, comparable accuracy is obtained using the \ac{AR}-based approach. In addition, it can be seen that even though the unsynchronized receives' performance improved by a factor of 3.4 in Table \ref{AVres2} w.r.t. Table \ref{AVres1}, synchronized ones improved by nearly a factor of 4.5, showing the significance of the clock offsets corrected using the proposed framework. Furthermore, the mean localization errors of the proposed framework presented in Table \ref{AVres1} and Table \ref{AVres2} shows a comparable performance to the 199\,m mean error reported in \cite{strohmeierKNN}. However, unlike our proposed method which utilizes both \ac{GSN} and \ac{SN} receivers, the \ac{TDoA}-based \ac{MLAT} method in \cite{strohmeierKNN} only considers the \ac{GPS}-synchronized \ac{GSN} receivers.

\begin{table}[t]
	\caption{Localization Error in Figure \ref{Res1}}
	\centering
	\begin{tabular}{ | l || c | c |}
		\hline
		\textbf{AV tracking method} & \textbf{Mean} $\mathcal{E}$ [m] & \textbf{Median} $\mathcal{E}$ [m]\\
		\hline
		\hline
		Without synchronization & 104$\times$10$^3$ & 60$\times$10$^3$\\
		\hline
		AR-based sync. without KF$_2$ &  951 &  540\\
		\hline
		AR-based synchronization &  719 & 460\\
		\hline
		LSTM-based sync. without KF$_2$ & 940 & 512 \\ 
		\hline
		LSTM-based synchronization & 680 & 420\\
		\hline
	\end{tabular}
	\label{AVres1}
\end{table}

\begin{table}[t]
	\caption{Localization Error in Figure \ref{Res2}}
	\centering
	\begin{tabular}{ | l || c | c |}
		\hline
		\textbf{AV tracking method} &\textbf{Mean} $\mathcal{E}$ [m] & \textbf{Median} $\mathcal{E}$ [m]\\
		\hline
		\hline
		Without synchronization & 30$\times$10$^3$ & 24$\times$10$^3$\\
		\hline
		AR-based sync. without KF$_2$ &  200 &  108\\
		\hline
		AR-based synchronization &  160 & 105\\
		\hline
		LSTM-based sync. without KF$_2$ & 210 & 106 \\ 
		\hline
		LSTM-based synchronization & 155 & 103\\
		\hline
	\end{tabular}
	\label{AVres2}
\end{table}

The performance analysis of the proposed framework illustrates that \ac{LSTM}-based and ARIMA-based methods provide comparable performances. In order to select one approach over the other, several aspects must be taken into account. The first aspect is the computational power since training LSTM models requires significantly more computational power than training ARIMA models. This article assumes that the model training is done at the network backend, which typically has sufficient computational power \cite{electrosense1}. However, if a CWN opts for model training to run on the receiver's companion computer, then the ARIMA model is a more suitable approach. Another important aspect is the model tuning and updating over time as LSTM models provide a more sustainable and futureproof solution compared to ARIMA models. For all sensors considered in this work, we used the same LSTM-based RNN architecture, showing promising synchronization capabilities. On the contrary, choosing the ARIMA model's orders requires special attention to the stationarity, the amount of correlation, and the considered training period in the time-series data, which all may vary from one receiver to another.

\section{Conclusion}\label{G2A:CONC}
The localization of \acp{AV} using a partially unsynchronized \ac{CWN} has been investigated. Particularly, a noncoherent \ac{CWN} with a mix of synchronized and unsynchronized receivers has been considered. This paper proposed a framework for the localization and tracking of \acp{AV}, including spoofing ones. The proposed framework investigated two synchronization approaches for the noncoherent \ac{CWN}: \ac{AR}-based and LSTM-based. Subsequently, we used a \ac{TDoA}-based \ac{MLAT}, along with a \ac{KF}, to estimate the location of the targeted \ac{AV}. The proposed method has been validated using an OpenSky dataset, where it proved that both synchronization approaches were able to provide significant gains in terms of \acp{AV} localization accuracy, improving it by orders of magnitude. Finally, the reported results motivate investigating other deep-learning-based methods, such as convolutional neural networks (CNNs), with CWNs, where datasets recorded over several hours could potentially enable end-to-end location estimation.

\balance
\bibliographystyle{ieeetr} 
\bibliography{IEEEosky}

\end{document}

%% file: Acro_G2A.tex
\begin{acronym}
\acro{A2A}{air-to-air}
\acro{A2G}{air-to-ground}
\acro{ADr}{amateur drone}
\acro{ADS-B}{automatic dependent surveillance-broadcast}
\acro{AN}{anchor}
\acro{AoA}{angle of arrival}
\acro{AR}{autoregressive}
\acro{ARIMA}{autoregressive integrated moving average}
\acro{ATM}{air traffic management}
\acro{ACF}{autocorrelation function}
\acro{AV}{aerial vehicle}

\acro{BS}{base station}

\acro{CA}{collision avoidance}
\acro{CDF}{cumulative distribution function}
\acro{CRLB}{Cram\'{e}r-Rao lower bound}
\acro{CWN}{crowdsourced wireless network}

\acro{D2D}{device-to-device}

\acro{G2A}{ground-to-air}
\acro{G2G}{ground-to-ground}
\acro{GPS}{global positioning system}
\acro{GSN}{GPS-enabled sensor}

\acro{IoT}{Internet of things}

\acro{KF}{Kalman filter}

\acro{LAN}{local area network}
\acro{LoS}{line-of-sight}
\acro{LPWAN}{low-power wide-area network}
\acro{LSTM}{long short-term memory}

\acro{MAV}{manned aerial vehicle}
\acro{ML}{machine learning}
\acro{MLAT}{multilateration}

\acro{NB}{narrowband}
\acro{NLoS}{non-line-of-sight}

\acro{PACF}{partial ACF}
\acro{PDF}{probability density function}
\acro{PL}{path loss}

\acro{RF}{radio frequency}
\acro{RndF}{random forest}
\acro{RNN}{recurrent neural network}
\acro{RSSI}{received signal strength indicator}
\acro{RSS}{received signal strength}
\acro{RTT}{round trip time}

\acro{SDR}{software-defined radio}
\acro{SN}{non-GPS sensor}
\acro{SNR}{signal-to-noise ratio}
\acro{SS}{synchronized sensors} 
\acro{SVM}{support vector machines}

\acro{TD}{Telecom Design}
\acro{TDoA}{time differences of arrival}
\acro{TN}{target node}
\acro{ToA}{time of arrival}
\acro{ToF}{time of flight}

\acro{UAV}{unmanned aerial vehicle}
\acro{UNB}{ultra narrowband}
\acro{US}{unsynchronized sensors}
\acro{UTM}{UAV traffic management}

\acro{WAN}{wide-area network}
\acro{WC}{well-clear}
\end{acronym}

%% file: clkErrModel.tex
\begin{tikzpicture}
	
\begin{axis}[%
legend style={font=\scriptsize},
width=7cm,
height=5.5cm,
scale only axis,
bar shift auto,
xmin=-450,
xmax=595.08,
xlabel={Clock offset [nsec]},
ymin=0,
ymax=900,
ylabel={Count},
axis background/.style={fill=white},
xmajorgrids,
ymajorgrids,
]
\addplot[ybar, bar width=22.6, fill=mycolor1, fill opacity=0.7, draw=myhist, area legend] table[row sep=crcr] {%
	-428.7	3\\
	-406.1	3\\
	-383.5	7\\
	-360.9	13\\
	-338.3	10\\
	-315.7	13\\
	-293.1	20\\
	-270.5	17\\
	-247.9	25\\
	-225.3	25\\
	-202.7	67\\
	-180.1	109\\
	-157.5	115\\
	-134.9	190\\
	-112.3	369\\
	-89.7	412\\
	-67.1	537\\
	-44.5	708\\
	-21.9	815\\
	0.700000000000017	826\\
	23.3	752\\
	45.9	585\\
	68.5000000000001	504\\
	91.1000000000001	508\\
	113.7	427\\
	136.3	389\\
	158.9	311\\
	181.5	316\\
	204.1	292\\
	226.7	307\\
	249.3	344\\
	271.9	364\\
	294.5	378\\
	317.1	334\\
	339.7	333\\
	362.3	265\\
	384.9	211\\
	407.5	152\\
	430.1	89\\
	452.7	65\\
	475.3	46\\
	497.9	39\\
	520.5	18\\
	543.1	14\\
	565.7	8\\
};
\addplot[forget plot, color=white!15!black] table[row sep=crcr] {%
	-458.08	0\\
	595.08	0\\
};

\addplot [color=mycolor2, line width=2.0pt]
table[row sep=crcr]{%
	-374.306228514463	8.18520152186642\\
	-364.984444622087	9.24462491155135\\
	-355.662660729711	10.2627671203109\\
	-346.340876837335	11.2017362950111\\
	-337.019092944959	12.0609530384911\\
	-327.697309052583	12.885661850917\\
	-318.375525160207	13.7419584634803\\
	-309.053741267831	14.6779009598911\\
	-299.731957375454	15.7105423870015\\
	-290.410173483078	16.8552159930664\\
	-281.088389590702	18.1696481317815\\
	-271.766605698326	19.7979322130726\\
	-262.44482180595	21.9678564806108\\
	-253.123037913574	24.9768425839729\\
	-243.801254021198	29.1992730070841\\
	-234.479470128822	35.025176168634\\
	-225.157686236446	42.7744342126455\\
	-215.83590234407	52.5209105431033\\
	-206.514118451694	64.0185249746972\\
	-197.192334559318	76.8049494016346\\
	-187.870550666942	90.5517470868925\\
	-178.548766774566	105.505187293967\\
	-169.226982882189	122.714533749528\\
	-159.905198989813	143.820191781648\\
	-150.583415097437	170.368323546754\\
	-141.261631205061	202.97883749152\\
	-131.939847312685	240.862533846575\\
	-122.618063420309	282.127724973356\\
	-113.296279527933	324.732668152704\\
	-103.974495635557	367.562628293702\\
	-94.6527117431808	410.915243286591\\
	-85.3309278508048	456.072455302971\\
	-76.0091439584287	504.242081849699\\
	-66.6873600660526	555.477223283546\\
	-57.3655761736765	608.139604400852\\
	-48.0437922813005	659.111817017999\\
	-38.7220083889244	704.589319561816\\
	-29.4002244965483	740.999135354646\\
	-20.0784406041722	765.683115977417\\
	-10.7566567117962	777.167442930132\\
	-1.43487281942004	775.138648113062\\
	7.88691107295602	760.335740823563\\
	17.2086949653321	734.567095226451\\
	26.5304788577081	700.680343605679\\
	35.8522627500842	662.293617305242\\
	45.1740466424603	623.211602739598\\
	54.4958305348363	586.617289406792\\
	63.8176144272125	554.382520320262\\
	73.1393983195886	526.770342657253\\
	82.4611822119646	502.742961918065\\
	91.7829661043407	480.670549312505\\
	101.104749996717	459.174608159033\\
	110.426533889093	437.655350883563\\
	119.748317781469	416.310893976153\\
	129.070101673845	395.823435731079\\
	138.391885566221	376.959090236756\\
	147.713669458597	360.28992245013\\
	157.035453350973	346.126967654569\\
	166.357237243349	334.542525395339\\
	175.679021135725	325.455607205712\\
	185.000805028101	318.750749385342\\
	194.322588920478	314.356595804659\\
	203.644372812854	312.306721446747\\
	212.96615670523	312.672391092932\\
	222.287940597606	315.489391744978\\
	231.609724489982	320.625103107841\\
	240.931508382358	327.663495578765\\
	250.253292274734	335.807468533238\\
	259.57507616711	343.949493791713\\
	268.896860059486	350.836835090295\\
	278.218643951862	355.358306406404\\
	287.540427844238	356.780369680853\\
	296.862211736614	354.839278170385\\
	306.18399562899	349.674787194953\\
	315.505779521366	341.60192203739\\
	324.827563413743	330.889210559448\\
	334.149347306119	317.66292365835\\
	343.471131198495	301.978882665088\\
	352.792915090871	283.960943282154\\
	362.114698983247	263.914363113313\\
	371.436482875623	242.307521479135\\
	380.758266767999	219.689589326712\\
	390.080050660375	196.639280668381\\
	399.401834552751	173.770491831599\\
	408.723618445127	151.717349117523\\
	418.045402337503	131.11628685252\\
	427.367186229879	112.497388091394\\
	436.688970122256	96.1770280913885\\
	446.010754014632	82.2214968222911\\
	455.332537907008	70.4662551785591\\
	464.654321799384	60.5987806480267\\
	473.97610569176	52.2444178048283\\
	483.297889584136	45.0353674696061\\
	492.619673476512	38.6477161731307\\
	501.941457368888	32.8521602567351\\
	511.263241261264	27.5620512581263\\
	520.58502515364	22.8031539733533\\
	529.906809046016	18.6505466513271\\
	539.228592938392	15.1246904471296\\
	548.550376830768	12.152978852071\\
};

\end{axis}
\end{tikzpicture}%

%% file: ACF.tex
%
%
%
\begin{tikzpicture}

\begin{axis}[%
legend style={font=\scriptsize},
width=7cm,
height=3cm,
scale only axis,
xmin=0,
xmax=20,
xlabel={Lag},
ymin=-0.6,
ymax=0.6,
ytick={-0.6,-0.3,0,0.3,0.6},
ylabel={Sample ACF},
axis background/.style={fill=white},
xmajorgrids,
ymajorgrids,
legend style={legend cell align=left, align=left, draw=white!15!black}
]
\addplot[ycomb, color=mycolor1, mark size=2.0pt, mark=*, mark options={solid, fill=mycolor1, mycolor1}] table[row sep=crcr] {%
0	1\\
1	-0.501148384207578\\
2	-0.0218386804996533\\
3	0.0118981933447382\\
4	0.0170196881914937\\
5	0.00381881312073386\\
6	-0.0243786730341856\\
7	0.026769566356923\\
8	-0.0110254160558123\\
9	-0.00422634284875866\\
10	0.00138077513615816\\
11	0.00856578554786085\\
12	-0.00788872966741737\\
13	-0.00194565340923395\\
14	0.000872762188976965\\
15	0.0114629794458677\\
16	-0.01177617587847\\
17	0.00446632542676266\\
18	-0.00616684136749712\\
19	0.0121048534894362\\
20	-0.018541502956416\\
};
\addplot[forget plot, color=white!15!black] table[row sep=crcr] {%
0	0\\
20	0\\
};
\addlegendentry{Measured offset}

\addplot [color=mycolor2, forget plot]
  table[row sep=crcr]{%
0.5	0.106528925698154\\
20	0.106528925698154\\
};
\addplot [color=mycolor2]
  table[row sep=crcr]{%
0.5	-0.106528925698154\\
20	-0.106528925698154\\
};
\addlegendentry{Confidence bounds}

\addplot [color=black, forget plot]
  table[row sep=crcr]{%
0	0\\
20	0\\
};
\end{axis}

\end{tikzpicture}%

%% file: PACF.tex
%
%
%
\begin{tikzpicture}

\begin{axis}[%
legend style={font=\scriptsize},
width=7cm,
height=3cm,
scale only axis,
xmin=0,
xmax=20,
xlabel={Lag},
ymin=-0.6,
ymax=0.6,
ytick={-0.6,-0.3,0,0.3,0.6},
ylabel={Sample PACF},
axis background/.style={fill=white},
xmajorgrids,
ymajorgrids,
legend style={legend cell align=left, align=left, draw=white!15!black}
]
\addplot[ycomb, color=mycolor1, mark size=2.0pt, mark=*, mark options={solid, fill=mycolor1, mycolor1}] table[row sep=crcr] {%
0	1\\
1	-0.501148384207578\\
2	-0.36454333340711\\
3	-0.28602566023854\\
4	-0.216663375100689\\
5	-0.162190431620407\\
6	-0.163412910479489\\
7	-0.118894390990391\\
8	-0.0842215898501\\
9	-0.0986234547644024\\
10	-0.092496765563535\\
11	-0.0725726124580366\\
12	-0.0698021304008207\\
13	-0.0692306954572703\\
14	-0.0699247596146911\\
15	-0.048521147441055\\
16	-0.0500003210482767\\
17	-0.0419001449106602\\
18	-0.0489293181337457\\
19	-0.0318557997813708\\
20	-0.0518023714458429\\
};
\addplot[forget plot, color=white!15!black] table[row sep=crcr] {%
0	0\\
20	0\\
};
\addlegendentry{Measured offset}

\addplot [color=mycolor2, forget plot]
  table[row sep=crcr]{%
0.5	0.106528925698154\\
20	0.106528925698154\\
};
\addplot [color=mycolor2]
  table[row sep=crcr]{%
0.5	-0.106528925698154\\
20	-0.106528925698154\\
};
\addlegendentry{Confidence bounds}

\addplot [color=black, forget plot]
  table[row sep=crcr]{%
0	0\\
20	0\\
};
\end{axis}
\end{tikzpicture}%

%% file: fittingResidualhistn_2.tex
\usepgfplotslibrary{
	groupplots,
}
\pgfplotsset{
	compat=1.3,
	my style/.style={
		width=4cm,height=4cm,
		xlabel={Normalized offset},
		ylabel=Count,
	},
	my legend style/.style={
		legend entries={
			Measured offset\,\,\,\,\,,
			LSTM residual\,\,\,\,\,,
			ARIMA residual,
		},
		legend style={
			at={([yshift=2pt]0,1)},
			anchor=south west,
			legend cell align=left, align=left, draw=white!15!black,
			font=\scriptsize,
		},
		legend columns=4,
	},
	cycle multiindex* list={
		mycolor1!75!black,
		mycolor2!75!black,
		mycolor3!75!black
		\nextlist
	},
}

\begin{tikzpicture}
	\begin{groupplot}[
		group style={
			group size=3 by 1,
			horizontal sep = 0.2cm,
			x descriptions at=edge bottom,
			y descriptions at=edge left,
		},
		my style,
		%
		xmin=-3.5,xmax=3.5,
		ymin=0,ymax=1600,
		xtick={-2,0,2},
		xmajorgrids,
		ymajorgrids,
		bar shift auto,
		]
		\nextgroupplot[
		my legend style,
		]
		\addplot[ybar, bar width=2.1, fill=mycolor1,opacity=0.7, draw=myhist, area legend] table[row sep=crcr] {%
			-3.3795	5\\
			-3.1385	13\\
			-2.8975	11\\
			-2.6565	19\\
			-2.4155	43\\
			-2.1745	72\\
			-1.9335	130\\
			-1.6925	234\\
			-1.4515	340\\
			-1.2105	514\\
			-0.9695	606\\
			-0.7285	784\\
			-0.4875	1305\\
			-0.2465	1530\\
			-0.0055	1266\\
			0.2355	1004\\
			0.4765	743\\
			0.7175	646\\
			0.9585	599\\
			1.1995	723\\
			1.4405	695\\
			1.6815	557\\
			1.9225	310\\
			2.1635	143\\
			2.4045	69\\
			2.6455	22\\
			2.8865	12\\
			3.1275	4\\
		};
		\addplot[ybar, bar width=2.3, fill=mycolor2,opacity=0.7, draw=myhist, area legend] table[row sep=crcr]{-1 -1\\};
		\addplot[ybar, bar width=2.3, fill=mycolor3,opacity=0.7, draw=myhist, area legend] table[row sep=crcr]{-1 -1\\};
		\nextgroupplot[
		cycle list shift=1,
		]
		\addplot[ybar, bar width=2.1, fill=mycolor2,opacity=0.7, draw=myhist, area legend] table[row sep=crcr] {%
			-3.6915	2\\
			-3.4745	1\\
			-3.2575	2\\
			-3.0405	8\\
			-2.8235	5\\
			-2.6065	9\\
			-2.3895	17\\
			-2.1725	39\\
			-1.9555	67\\
			-1.7385	125\\
			-1.5215	123\\
			-1.3045	239\\
			-1.0875	297\\
			-0.8705	451\\
			-0.6535	642\\
			-0.4365	823\\
			-0.2195	931\\
			-0.0025	1091\\
			0.2145	1178\\
			0.4314	1124\\
			0.6485	1024\\
			0.8655	919\\
			1.0825	783\\
			1.2995	663\\
			1.5165	515\\
			1.7335	432\\
			1.9505	326\\
			2.1675	236\\
			2.3845	167\\
			2.6015	99\\
			2.8185	53\\
			3.0355	41\\
			3.2525	21\\
			3.4695	20\\
			3.6865	12\\
			3.9035	4\\
			4.1205	6\\
			4.3375	2\\
			4.5545	1\\
			4.7715	1\\
		};
		\nextgroupplot[
		cycle list shift=2,
		]
		\addplot[ybar, bar width=2.8, fill=mycolor3,opacity=0.7, draw=myhist, area legend] table[row sep=crcr] {%
			-5.255	3\\
			-4.965	4\\
			-4.675	1\\
			-4.385	3\\
			-4.095	5\\
			-3.805	5\\
			-3.515	18\\
			-3.225	19\\
			-2.935	34\\
			-2.645	59\\
			-2.355	107\\
			-2.065	182\\
			-1.775	253\\
			-1.485	380\\
			-1.195	558\\
			-0.905000000000001	844\\
			-0.615000000000001	1154\\
			-0.325000000000001	1495\\
			-0.0350000000000006	1568\\
			0.254999999999999	1571\\
			0.544999999999999	1370\\
			0.835	1069\\
			1.125	690\\
			1.415	498\\
			1.705	296\\
			1.995	133\\
			2.285	73\\
			2.575	48\\
			2.865	22\\
			3.155	11\\
			3.445	6\\
			3.735	3\\
			4.025	5\\
			4.315	5\\
			4.605	2\\
			4.895	1\\
			5.185	2\\
			5.475	0\\
			5.765	1\\
			6.055	1\\
		};
	\end{groupplot}
\end{tikzpicture}